\documentstyle[12pt]{article}
\begin{document}
\def\om{\omega}
\def\omt{\tilde{\omega}}
\def\ti{\tilde}
\def\o{\Omega}
\def\bchi{\bar\chi^i}

\def\ba{\bar a}
\def\w{\wedge}

\def\Tr{{\rm Tr}}
\def\ST{{\rm STr}}
\def\ss{\subset}
\def\bc{{\bf C}}
\def\br{{\bf R}}

\def\al{\alpha}
\def\la{\langle}
\def\ra{\rangle}

\def\th{\theta}
\def\lm{\lambda}

\def\d{\partial}
\def\dz{\partial_z}
\def\dbz{\partial_{\bar z}}

\def\be{\begin{equation}}
\def\ee{\end{equation}}
\def\bea{\begin{eqnarray}}
\def\eea{\end{eqnarray}}
\def\A{{\cal A}}
\def\B{{\cal B}}
\def\T{{\cal T}}
\def\bT{\bar{\cal T}}
\def\Z{{\cal Z}}
\def\si{\sigma}
\def\*{\ddagger}
\def\j{\dagger}
\def\bz{\bar{z}}
\def\e{\varepsilon}
\def\b{\beta}
\def\bb{\bar b}
\def\bk{\bar k}
\begin{titlepage}
\begin{flushright}
{}~
hep-th/9809166

IML/98-25
\end{flushright}

\vspace{3cm}
\begin{center}
{\Large \bf Noncommutative integrability}\\ [50pt]{\small
{\bf C. Klim\v{c}\'{\i}k}\\Institute de Math\'ematiques de Luminy\\163,
Avenue de Luminy, 13288 Marseille, France}

\vskip2pc

\vspace{1cm}
\begin{abstract}
I compute the cohomology of a non-commutative complex underlying the notion
of the gauge field on the fuzzy sphere.
\vskip2pc

\end{abstract}
\end{center}

\end{titlepage}

\noindent 1. Noncommutative geometry is a well established mathematical
discipline
with a surprising and nontrivial impact on quantum field theory in general
and the standard model in particular \cite{Con}. There is a subactivity
in that vast subject which aims to replace the field theoretical models
on the standard smooth manifolds by its counterparts defined on suitable
noncommutative deformations of those manifolds \cite{GKP,Ma,Hop}.
The resulting noncommutative models usually respect all symmetries
or supersymmetries of the commutative theories but they have an important
advantage of possessing only a finite (though large) number of degrees
of freedom. Recently such structures have emerged also in the context of
the matrix model of
$M$-theory \cite{BFSS,IKKT,CDS}.

The basic idea of the approach is as follows: One considers an Euclidean
space-time which is taken to be compact for convenience. This spacetime
gets equipped with a symplectic structure. A quantization of this symplectic
structure gives an algebra of quantum observables which is to be taken
as the {\it definition} of the non-commutative manifold. The compactness
results in the finiteness of that noncommutative algebra of observables.
The important feature of the formalism is that the Hamiltonian vector
fields on the classical manifold survive the deformation. They are generated
by the quantized Hamiltonians via the commutators. Finally, also the
integration
over the symplectic manifolds gets replaced in the deformed picture by the
operation of taking the trace over the matrix algebra.

Having at hand the deformed notions of algebra, Hamiltonian vector fields
and integration we can construct the field theoretical actions for the
models involving the scalar fields  on the deformed noncommutative
manifold. As an example, consider  a Riemann sphere as a spacetime of an
Euclidean field theory.

 The crucial observation is that
$S^2$ is naturally a symplectic manifold; the symplectic form $\omega$ is
up to a normalization
just
the round
volume form on the sphere. Using the standard complex coordinate $z$ on
the Riemann sphere, we have
$$\omega=-{N\over 2\pi}{d\bz \w dz\over (1+\bz z)^2},\eqno(i)$$ with $N$
a real
 parameter\footnote{Note, that we have chosen a normalization which makes
the
form $\omega$ purely imaginary. Under quantization, hence, the Poisson bracket
is replaced
 by a commutator {\it without} any imaginary unit factor.}. If we consider
 a scalar field theory,
then the scalar field $\phi$ is a function on the symplectic manifold or,
in other words, a classical
observable. The action of the massless (real) scalar field theory on $S^2$ is
 given by $$S=-i\int \omega R_i\phi R_i\phi,\eqno(ii)$$ where $R_i$ are the
 vector fields
 which generate the $SO(3)$ rotations of $S^2$ and the Einstein summation
convention is understood.
The vector fields $R_i$ are Hamiltonian; this means that there exists three
concrete observables
$r_i$ such that $$ \{r_i,\phi\}=R_i\phi.\eqno(iii)$$
Here $\{.,.\}$ is the Poisson bracket which corresponds to the symplectic
 structure $\omega$.
 The observables $r_i\in{\br}^3$ are just the coordinates of the embedding
of $S^2$ in $\br^3$.
Thus we can rewrite the action $(ii)$ as
$$S=-i\int\omega\{r_i,\phi\}\{r_i,\phi\}.\eqno(iv)$$

Suppose we quantize the symplectic structure on $S^2$ (probably the first who
 has done
 it was Berezin \cite{Ber}). Then the algebra of observables becomes the
noncommutative algebra
of all square matrices with entries in $\bc$; the quantization of $S^2$
 can be only
performed if $N$ is an integer,
the size of the scalar field matrices $\phi$ is then $(N+1)\times (N+1)$.
 This algebra of matrices
defines the noncommutative (or fuzzy \cite{Ma}) sphere.
The integration over the
phase space volume
form $i\omega$ is replaced by
taking a properly normalized trace $\Tr$ over the matrices and the Poisson
 brackets are replaced by
commutators (the Hamiltonians $r_i$
are also quantized, of course).

Putting together, we can consider along with $(iv)$ a noncommutative action
$$S=-{1\over N+1}\Tr([r_i,\phi],[r_i,\phi]).\eqno(v)$$

The action $(v)$ has a few nonstandard properties. First of all, the space
of all "fields"(=matrices)
is finite dimensional and the product of fields is noncommutative. The latter
 property may seem
awkward but in all stages of analysis we shall never encounter a problem
 which this noncommutativity
might create. The former property, however, is highly desirable, since
all divergences of the usual
field theories are automatically eliminated. We may interpret $(v)$ as
 the regularized version of $(iv)$;
 the fact that $(v)$ goes to $(iv)$ in the limit $N\to\infty$ is just
the statement that classical mechanics
 is the limit of the quantum one for the value of the Planck constant $1/N$
approaching zero. Remarkably,
unlike lattice regularizations, $(v)$ preserves the  $SO(3)$ isometry
of the sphere (="spacetime"). Indeed,
 under the variation $\delta\phi=[r_i,\phi]$ the action $(v)$ remains
 invariant.
\vskip1pc
\noindent 2. A question of obvious interest consists in enlarging
the above-mentioned framework of constructing
the field theories on noncommutative manifolds also to the case of
 nonscalar fields. In practice, one is interested
in spinor and 1-form fields (gauge potentials). While the quantization
 gives automatically
 scalars the rest
of the story is not evident because in the literature on quantization
 one did not consider the question of quantizing
the vector bundles as the moduls of algebra. The idea adopted in
\cite{GKP} for quantizing spinors
is simple: one enlarges the algebra appropriately to include
the spinor fields with  the scalars. The resulting
enlarged algebra is known as the algebra of superfields; by the way,
 one gains in this way a possibility
of constructing noncommutative  supersymmetric field theories along
the same lines as above. The issue of the
gauge fields turned out to be more complicated than the story of spinors .
 It was necessary to construct
the deformation of the whole de Rham complex for being able to define
the notion of the gauge field in the
noncommutative case \cite{K}.

Actually, one needs more than the deformation of the algebraic
structure of the de Rham complex. Indeed, the
notion
of the exterior derivative $d$ has to survive the quantization.  Since $d$
 has to be a derivative, one has to
find Hamiltonian vector fields to express $d$. It turned out that this can
be done by paying the price of injecting
the standard classical de Rham complex to a larger complex which can be
 deformed with all its relevant structures.
This new complex was referred to as the Hamiltonian de Rham complex in
 \cite{K}, reflecting the fact that
the exterior derivative could be expressed in terms of the Hamiltonian
 vector fields and thus quantized.
There remains a mathematical\footnote{This question can well become physical
  in the context of the so-called
world-sheet
$T$-duality.}
 question which was only touched upon in \cite{K} but which is quite
important in order to have  a feeling
of general consistency of the deformed picture. The question reads:
What is the cohomology of the deformed
complex? A satisfactory answer must be that it is the same as the
cohomology of the undeformed complex.
The reason for this is simple: the quantization should influence only
the short distance properties of the
manifold but not its topology; obviously, the cohomology of the de Rham
complex reflects the topology
of the underlying manifold. In this note, we give a so far missing proof
that the deformation of the sphere
does not change the cohomological content of the Hamiltonian de Rham complex.

\vskip1pc
\noindent 3. We should first review what are the nondeformed and deformed
 Hamiltonian complexes over the
sphere following \cite{K}, then we shall actually compute their cohomologies.
 We shall not review the way how the
standard de Rham complex is injected into the Hamiltonian one.
The interested reader may find it again in \cite{K}.

Consider the algebra of functions on the complex $C^{2,1}$ superplane,
i.e. algebra
generated by bosonic variables $\bar\chi^i,\chi^i, i=1,2$ and by
fermionic ones $\bar a,a$.
The algebra is equipped with the graded involution \be (\chi^i)^{\*}=
\bar\chi^i,\quad, (\bar\chi^i)^{\*}=\chi^i,\quad, a^{\*}=\bar a,
\quad, \bar a^{\*}=-a\ee
and with the super-Poisson bracket
\be \{f,g\}=
\d_{\chi^i}f\d_{\bar\chi^i}g-\d_{\bar\chi^i}f\d_{\chi^i}g +
(-1)^{f+1}[\d_a f\d_{\bar a}g+\d_{\bar a} f\d_a g].\ee
Here and in what follows, the Einstein summation convention applies.
We can now apply the (super)symplectic reduction with respect
to a moment map $\bar\chi^i\chi^i +\bar a a $. The result is a
smaller algebra
$\A$, that by definition consists of all functions $f$ with
the property
\be \{f,\bar\chi^i\chi^i+\bar a a\}=0.\ee Moreover, two functions
obeying (55)
are considered to be equivalent if they differ just by a product
of $(\bar\chi^i\chi^i+\bar a a-1)$ with some other such function.
The algebra $\A$ has a subalgebra $\A_e$ which consists of all even
 elements of $\A$; that means
 that the odd generators
$a, \bar a$ appears only in the combination $\bar aa$. We identify
$\A_e$ with the space of the
 (complex) Hamiltonian
$0$-forms $\Omega_0$ and also with the space of the (complex)
 Hamiltonian $2$-forms $\Omega_2$.
The space of the (complex) Hamiltonian
$1$-forms is defined as
\be \Omega_1\equiv \A_{a}\oplus\A_{a}\oplus\A_{\ba }\oplus\A_{\ba },\ee
where the space $\A_a$ ($\A_{\ba}$) consists of all odd elements of
 $\A$ not depending on $\ba$($a$).

In order to define the exterior derivative $d$ we introduce
the following Hamiltonian vector fields
$T_i,\bar T_i$
\be T_i=\bar\chi^i\d_{\bar a}-a\d_{\chi^i},\quad \bar T_i=
\bar a\d_{\bar\chi^i} +\chi^i\d_a.\ee
Of course, they annihilate the moment map
$(\bar\chi^i\chi^i +\bar a a)$, otherwise
they would not be well defined differential operators acting on $\A$.
Their Hamiltonians are
\be t_i =\bar\chi^i a,\quad \bar t_i=\chi^i \bar a.\ee

 The multiplication in $\o$ is entailed by one in $\A$,
the only non-obvious thing is to define the product of 1-forms.
 Here it is
\be (A_1,A_2,\bar A_1,\bar A_2) (B_1,B_2,\bar B_1,\bar B_2)
\equiv A_1 \bar B_1 +A_2 \bar B_2 +\bar A_1 B_1 +
\bar A_2 B_2.\ee Of course, the r.h.s. is viewed as an element of $\o_2$.
The product of a 1-form and a 2-form is set to zero by definition.
Now the coboundary operator
 $d$ is given by \be df\equiv (T_1 f, T_2 f, \bar T_1 f, \bar T_2 f),
f\in \o_0;\ee
\be d(A_1,A_2,\bar A_1,\bar A_2)
\equiv T_1 \bar A_1 +T_2 \bar A_2 +\bar T_1 A_1 +\bar T_2 A_2,
\quad (A_1,A_2,\bar A_1,\bar A_2)\in \o_1;\ee \be dh=0, \quad h\in \o_2.\ee
It maps $\o_i$ to $\o_{i+1}$ and it satisfies \be d^2=0, \quad d(AB)=
(dA)B +(-1)^A A(dB).\ee

There remains to clarify the issues of reality and cohomology. An involution
 $\j$ is defined as follows
\be f^{\j}=f^{\*}, ~f\in \o_0, \quad h^{\j}=-h^{\*},~h\in\o_2;\ee
\be (A_i,\bar A_i)^{\j}=
(\bar A_i^{\*},-A_i^{\*}), ~(A_i,\bar A_i)\in \o_1.\ee The involution
$\j$ ($\j^2=1$) preserves
the linear combinations with real coefficients and the multiplication,
and commutes with the coboundary operator $d$:
\be (af+bg)^{\j}=af^{\j} +bg^{\j}, ~a,b\in{\bf R},~f,g\in\o;\ee
\be (fg)^{\j}=f^{\j}g^{\j},~f,g\in\o;\ee
\be (df)^{\j}=df^{\j}, f\in\o.\ee
The real forms under the involution $\j$ form the real Hamiltonian complex.
 Its cohomology contains only two nontrivial classes: a $0$-form $1$ and
a $2$-form $i\ba a$.

Now we are ready to quantize the infinitely dimensional algebra $\A$ with
the goal of obtaining
its (noncommutative) finite dimensional deformation.
We start with the well-known quantization of the complex plane $C^{2,1}$.
The generators $\bchi,\chi^i,\ba, a$ become creation and annihilation
operators
on the Fock space whose commutation relations are given by the standard
replacement
\be \{.,.\}\to {1\over h}[.,.].\ee
Here $h$ is a real parameter (we have absorbed the imaginary unit into
the definition
of the Poisson bracket) referred to as the "Planck constant". Explicitely
\be [\chi^i,\bar\chi^j]_-=h\delta^{ij}, \quad [a,\bar a]_+=h\ee and all
remaining graded
commutators vanish. The Fock space is built up as usual, applying the
creation operators
$\bchi,\ba$ on the vacuum $\vert 0 \rangle$, which is in turn annihilated
 by the annihilation operators $\chi,a$. The scalar product on the Fock
space
is fixed by the requirement that the barred generators are adjoint of
the unbarred ones.
 We hope that using the same symbols for the classical and quantum generators
will not confuse the reader; it should be fairly obvious from the context
which usage we have in mind.

Now we perform the quantum symplectic reduction
with the self-adjoint moment map
$ (\bchi\chi^i +\ba a)$. First we restrict the Hilbert space only to the
 vectors
$\psi$ satisfying the constraint \be (\bchi\chi^i +\ba a -1)\psi=0.\ee
Hence operators $\hat f$ acting on this restricted space have to fulfil
 \be [\hat f,\bchi\chi^i +\ba a]=0\ee
and they are to form our deformed version of $\A$.

The spectrum of the operator $(\bchi\chi^i+\ba a -1)$ in the Fock space is
 given by a sequence $mh-1$, where $m$'s are integers. In order to
 fulfil (19)
for a non-vanishing $\psi$, we observe that the inverse Planck constant
$1/h$ must be an integer $N$. The constraint (19) then selects only
$\psi$'s living
in the eigenspace $H_N$
of the operator $(\bchi\chi^i +\ba a -1)$ with the eigenvalue $0$.
This subspace of the Fock space has the dimension $2N+1$ and the
algebra $\A_N$ of operators $\hat f$ acting on it is $(2N+1)^2$-dimensional.

When
$N\to\infty$ (the dimension $(2N+1)^2$ then also diverges) we have
the Planck constant approaching $0$ and, hence, the algebras $\A_N$ tend
to the classical limit $\A$.

The Hilbert space $H_N$ is naturally graded. The even subspace $H_{eN}$
is created
 from the Fock vacuum by applying only the bosonic creation operators:
\be (\bar\chi^1)^{n_1}(\bar\chi^2)^{n_2}\vert 0\rangle,\quad n_1+n_2=N,\ee
while
the odd one $H_{oN}$ by applying both bosonic and fermionic creation
operators:
\be (\bar\chi^1)^{n_1}(\bar\chi^2)^{n_2} \ba\vert 0\rangle, \quad n_1+n_2
=N-1.\ee
Correspondingly, the algebra of operators $\A_N$ on $H_N$ consists of
an even part $\A_{eN}$
(operators respecting the grading) and an odd part (operators reversing
the grading).
 The odd part can be itself written as a direct sum
$\A_{aN}\oplus\A_{\bar a N}$.
The two components in the sum are distinguished by their images:
$\A_{aN}H_N=H_{eN}$
while $\A_{\bar aN}H_N=H_{oN}$.  $\A_{aN}$ is spaned by operators
\be (\bar\chi^1)^{n_1}(\bar\chi^2)^{n_2}(\chi^1)^{m_1}(\chi^2)^{m_2}a,
\quad n_1+n_2=m_1+m_2 +1=N,\ee
$\A_{\bar aN}$ by
\be (\bar\chi^1)^{n_1}(\bar\chi^2)^{n_2}\ba (\chi^1)^{m_1}(\chi^2)^{m_2},
\quad n_1+n_2+1=m_1+m_2=N\ee
and $\A_{eN}$ by
\be (\bar\chi^1)^{n_1}(\bar\chi^2)^{n_2}(\chi^1)^{m_1}(\chi^2)^{m_2}
(\ba a)^k,\quad
n_1+n_2=m_1+m_2 =N-k.\ee

Here the graded involution $\*$ in the noncommutative algebra $\A_N$ is
defined exactly as in (1).

Define a non-commutative Hamiltonian de Rham complex $\o_N$ of the
fuzzy sphere $S^2$ as
the graded associative algebra with unit \be \o_N=\o_{0N} \oplus \o_{1N}
\oplus \o_{2N},\ee
where
\be \o_{0N}=\o_{2N}=\A_{eN}\ee
and
\be \o_{1N}=\A_{aN}\oplus\A_{aN}\oplus\A_{\ba N}\oplus\A_{\ba N}.\ee The
multiplication in $\o_N$
with the standard properties with respect to the grading is entailed by
one in $\A_N$.
 The product of 1-forms
is given by the same formula as in the
graded commutative case
\be (A_1,A_2,\bar A_1,\bar A_2) (B_1,B_2,\bar B_1,\bar B_2)\equiv A_1
\bar B_1 +A_2 \bar B_2 +
\bar A_1 B_1 +\bar A_2 B_2.\ee Of course, by definition, the r.h.s. is
viewed as an element
 of $\o_{2N}$. Here we note an important difference with the graded
commutative case:
the product $AA$ of a 1-form $A$ with itself automatically vanishes
in the commutative case
 but may be a non-vanishing element of $\o_{2N}$ in the deformed picture.
The product of a 1-form and a 2-form is again set to zero by definition.
 Now the coboundary
operator $d$ is given by \be df\equiv (T_1 f, T_2 f, \bar T_1 f, \bar T_2 f),
\quad f\in \o_{0N};\ee
\be d(A_1,A_2,\bar A_1,\bar A_2)\equiv T_1 \bar A_1 +T_2 \bar A_2 +
\bar T_1 A_1 +\bar T_2 A_2,
\quad (A_1,A_2,\bar A_1,\bar A_2)\in \o_{1N};\ee \be dh=0,
\quad h\in \o_{2N},\ee
where the action of $T_i,\bar T_i$ is given by the noncommutative
version of (8):
\be T_i X\equiv N(t_i X -(-1)^X X t_i), \quad \bar T_i X\equiv
N(\bar t_i X -(-1)^X X \bar t_i), \quad X\in \A_N,\ee where
\be \quad
t_i =\bar\chi^i a,\quad \bar t_i=\chi^i \bar a.\ee $d$ maps $\o_{iN}$ to
$\o_{i+1,N}$ and it satisfies
\be d^2=0, \quad d(AB)= (dA)B +(-1)^A A(dB).\ee Using the graded
involution $\*$, we define
the standard involution $\j( \j^2=1)$ on the noncommutative complex $\o_N$:
\be f^{\j}=f^{\*}, ~f\in \o_{0N}, \quad g^{\j}=-g^{\*},~g\in\o_{2N};\ee
\be (A_i,\bar A_i)^{\j}=
(\bar A_i^{\*},-A_i^{\*}), ~(A_i,\bar A_i)\in \o_{1N}. \ee
The coboundary map $d$ is compatible with the involution, however,
due to noncommutativity,
it is no longer true that the product of two real elements of $\o_N$ gives
a real element.
Thus we cannot define the real noncommutative Hamiltonian de Rham complex.
 For field theoretical
 applications this is not a drawback, nevertheless, because for the
formulation of the field theories
 we shall not need the structure of the real subcomplex, but only the
involution on
the complex Hamiltonian de Rham complex.

\vskip1pc

\noindent 5. We are now ready to compute the cohomology of the noncommutative
 complex $\o_N$.
\vskip1pc
\noindent {\bf Theorem}:

\noindent i) Let $f\in\o_{0N}, \quad df=0$. Then $f$ is the unit element of
$\o_{0N}$ (unit matrix acting on $H_N$) multiplied by some  number.

\noindent ii) Let $A\equiv(A_1,A_2,\bar A_1,\bar A_2)\in\o_{1N}, \quad dA=0,
\quad A=A^{\j}$.
Then $A=dg$ for some $g\in\o_{0N},~g=g^{\j}$.

\noindent iii) Let $F\in\o_{2N}$ (i.e. $dF$ automatically vanishes), $F=F^{\j}$.
 Then
$F$ can be written as $F=pId+dB$, where $B\in\o_{1N}, ~B=B^{\j}$ is some 1-form,
$Id$
is the unit element in $\A_{eN}$ and $p$ is an imaginary number. $Id\in\o_{2N}$
 itself
cannot be written as a coboundary of some 1-form.

\vskip1pc
Thus the theorem implies that
$Id$ is the only nontrivial cohomology class in $H^2(\o_N)$ and $H^0(\o_N)$,
and $H^1(\o_N)$
vanishes.
\vskip1pc
\noindent {\bf Proof}:

\noindent i) One notices that the Hamiltonians $t_i,\bar t_i$ of the vector
 fields $T_i,\bar T_i$
generates the whole algebra $\A_N$ and therefore also its subalgebra
$\A_{eN}=\o_{0N}$.
According to (33), the $T_i,\bar T_i$ act on an element $f\in\o_{0N}$
as commutators $N[t_i,f], N[\bar t_i,f]$, respectively.
Thus vanishing of the commutators means that $f$ commutes with all
matrices in $\A_N$.
Hence $f$ is a multiple of the unit matrix $Id$.

\noindent ii)
We want to show that $A_i=[t_i,\Phi]; \bar A_i=[\bar t_i,\Phi]$ for some
hermitian matrix $\Phi=\Phi^{\j};
\Phi\in\A_{0N}$.  The first step is to prove the following

\noindent {\bf Lemma}: Any real $1$-form $(A_i,\bar A_i)$ can be written
in terms of two hermitian
matrices $\Phi_1, \Phi_2$ which have zeros on their diagonals and two
 supertraceless
diagonal hermitian matrices $\Delta,
\tilde\Delta$ as follows
\bea A_1=[t_1, \Phi_1+\Delta +i\tilde\Delta] ; A_2=[t_2, \Phi_2+\Delta +
i\tilde\Delta];\cr \bar A_1=[\bar t_1,
\Phi_1+\Delta -i\tilde\Delta]; \bar A_2=[\bar t_2, \Phi_2+\Delta
-i\tilde\Delta].\eea

It easy to prove lemma by giving an explicit formula how to find $\Phi_i$
in terms of $A_i$. Here
it is in terms of the matrix elements
\be (\Phi_1)_{ij}={1\over i-j}(\{t_1,\bar A_1\}+\{\bar t_1, A_1\})_{ij},
(\Phi_2)_{ij}={1\over j-i}(\{t_2,\bar A_2\}+\{\bar t_2, A_2\})_{ij}; i\neq j.\ee
The formula giving the $\Delta$-s in terms of $A_i$ is somewhat cumbersome
and we invite the interested
reader to work it out as an exercise.

The proof of the part ii) of the theorem then finishes by noting that the
condition
 $dA=\{t_1,\bar A_1\}+\{\bar t_1, A_1\}+\{t_2,\bar A_2\}+\{\bar t_2, A_2\}=0$
clearly entails $\Phi_1=\Phi_2$
and one can also show that it gives $\tilde\Delta=0$.

\noindent iii) We have to show that every supertraceless antihermitian matrix
$\Psi=-\Psi^{\j}$
can be written as $\Psi=dA=\{t_1,\bar A_1\}+\{\bar t_1, A_1\}+\{t_2,\bar A_2\}
+\{\bar t_2, A_2\}$
for a certain real $1$-form $A=A^{\j}$. First of all we note that $dA$ is
always supertraceless; this
explains why $\Phi$ has to be supertraceless. Then we have to find for each
supertraceless antihermitian
$\Phi$ a set of   two hermitian
matrices $\Phi_1, \Phi_2$ which have zeros on their diagonals and two
supertraceless
diagonal hermitian matrices $\Delta,
\tilde\Delta$. Remind that according to ii) those data encode unambiguosly
a real one-form $A$.
One finds $\Delta=0$; to find  $\tilde\Delta$ is easy but the explicit formula
is somewhat cumbersome and
I do not list it here. There is slightly more work  needed to identify
$\Phi_1$
and $ \Phi_2$, on the other
hand the explicit formulas for the matrix elements of   $\Phi_1$ and  $\Phi_2$
are much nicer than those for
$\Delta$-s. Here there are:
\be (\Phi_1)_{ij}={1\over i-j}(\Psi)_{ij}; \quad (\Phi_2)_{ij}=
{1\over j-i}(\Psi)_{ij}.\ee
Of course this solution is not unique  for the "gauge" transformed form
$A +d\Phi$ gives also a solution.

 The only
nontrivial
cohomology class in $H^2(\o_N)$ is therefore supertraceful and can be chosen
to be an imaginary multiple of $Id$. The theorem is proved.
\vskip1pc
\noindent 6. We conclude by interpreting the result and scetching its
 possible application. The fact
that a closed one-form can be written as an exterior derivative of a
zero-form is often referred to by
saying that the one-form is integrable. We have shown here that this
integrability is not touched upon
by the noncommutative deformation of the complex. This result was by
no means evident and one  had to
use different technical tools than the infinitesimal calculus in order
to reveal the cohomological
content of the deformed complex. An interesting application of the result
 may reside in the world
of the string-theoretical target space duality. The sigma-models formulated
on the noncommutative
world-sheet become  dualizable in the similar way than their commutative
counterparts. The details of the
story are currently in preparation.

 \end{document}